\documentstyle[12pt,epsf]{article}
\newcommand{\be}{\begin{equation}}
\newcommand{\ee}{\end{equation}}
\newcommand{\bea}{\begin{eqnarray}}
\newcommand{\eea}{\end{eqnarray}}
\newcommand{\bt}{\begin{tabular}}
\newcommand{\et}{\end{tabular}}
\newcommand{\ba}{\begin{array}}
\newcommand{\ea}{\end{array}}

\begin{document}
\setcounter{page}{0}
\thispagestyle{empty}
\baselineskip=20pt
$~$
\hfill{
\begin{tabular}{l}
DSF$-$94/59\\
INFN$-$NA$-$IV$-$94/59
\end{tabular}}

\vspace{1.5cm}

\begin{center}
\begin{Large}
{\bf A Geometric Picture for Fermion Masses}
\end{Large}
\end{center}

\vspace{1.5cm}

\begin{center}
{\large
Salvatore Esposito$^{1}$ and Pietro Santorelli$^{2,1}$}
\end{center}

\vspace{0.5truecm}

\begin{center}
{\it
\noindent
$^{1}$INFN, Sezione di Napoli,
Mostra d'Oltremare Pad. 20, I-80125 Napoli Italy\\
\noindent
$^{2}$Dipartimento di Scienze Fisiche, Universit\`a di Napoli,
Mostra d'Oltremare Pad. 19, I-80125 Napoli Italy}
\end{center}

\vspace{1.5cm}

\begin{abstract}

\noindent
We describe a geometric picture for the pattern of fermion masses 
of the three generations which is invariant with respect to the 
renormalization group below the electroweak scale.
Moreover, we predict the upper 
limit for the ratio between the Dirac masses of the $\mu$ and $\tau$ 
neutrinos, $\left(m_{\nu_{\mu}}/ m_{\nu_{\tau}}\right) \leq \,
(9.6\,\pm\,0.6)\,\times\,10^{-3}$.

\end{abstract}

\newpage

The Standard Model \cite{GWS,W} of strong and electroweak interactions 
describes accurately all the observed phenomena of elementary particles.
In this framework, the fermion masses are essentially free parameters that 
must be fixed by experiment. 

Many efforts have been devoted to find 
relations among fermion masses, empirically and/or theoretically.
For example, a composite model for ``fundamental'' fermions proposed in 
\cite{TY} predicts the following relation among the masses of the three 
generations of quarks and leptons
\be
m_{f^{(3)}} \,\sim\,\sqrt{\frac{m_{f^{(2)}}^3}{m_{f^{(1)}}}}
\ee
which is experimentally well verified only for charged leptons.

Semi-empirical sum rules among masses can also be found,
as the so-called ``generation-changing mass-ratio sum rules'' \cite{T}

\bea
\sqrt{\frac{m_c}{m_u}} - \sqrt{\frac{m_s}{m_d}}  & = & 
              \sqrt{\frac{m_{\mu}}{m_e}} \nonumber\\
\sqrt{\frac{m_t}{m_c}} - \sqrt{\frac{m_b}{m_s}}  & = & 
              \sqrt{\frac{m_{\tau}}{m_{\mu}}} 
\eea
which works fairly well for the experimental values given in \cite{PDB}.
Furthermore, the previous formulas may indicate the quarks and leptons 
in the second and third generations as the excited states of the 
corresponding ones of the first generation.
Other relations among fermion masses, more or less elegant 
and experimentally verified have been proposed; see for example
\cite{list}.

Koide \cite{Koide} has shown that the last experimental data about 
the tau lepton mass \cite{PDB} are in excellent agreement with 
the following lepton mass formula
\be
m_e + m_{\mu} + m_{\tau} = \frac{2}{3}
   \left(\sqrt{m_e}+\sqrt{m_{\mu}}+\sqrt{m_{\tau}}\right)^2
\label{e:koide}
\ee
which he deduced in several different models \cite{Koide-models}.\\
An intriguing geometric interpretation of the previous 
formula has been recently proposed by Foot \cite{Foot}. 
He considers a three-dimensional Euclidean vector space and
the vector {\bf M} of components $(\sqrt{m_0}, \sqrt{m_0}, \sqrt{m_0})$ in 
a monometric, orthogonal, Cartesian reference frame and a vector {\bf L},
associated to the charged leptons, with components 
$(\sqrt{m_e}, \sqrt{m_{\mu}}, \sqrt{m_{\tau}})$. Here we generalize the 
suggestion of Foot which considers {\bf M}$=(1,1,1)$. Though the 
dimensional parameter $m_0$ does not enter explicitly in any physical 
expression which will be studied in the following, it may be viewed, 
for example, as the fundamental mass vector which generates by means of 
some unknown mechanism all known mass vectors.\\
By defining the angle $\vartheta_{ML}$ between these two vectors:
\be
\cos\vartheta_{ML} = \frac{ \mbox{{\bf M}} \cdot \mbox{{\bf L}} }
{|\mbox{{\bf M}}|\,
|\mbox{{\bf L}}|} =
\frac{1}{\sqrt{3}} \,
\frac{\sqrt{m_e} + \sqrt{m_{\mu}} + \sqrt{m_{\tau}}}
{\sqrt{m_e + m_{\mu} + m_{\tau}}} 
\label{e:foot}
\ee
\noindent
the Koide's mass formula (\ref{e:koide}) is reproduced for 
$\vartheta_{ML}=45^{\circ}$.
From the equation (\ref{e:foot}) and the experimental values of the 
masses of charged leptons \cite{PDB},
\be
\ba{lclcll}
m_e  & = & 0.51099906 & \pm & 0.00000015 & MeV\\
m_{\mu}  & = & 105.658389 &  \pm & 0.000034 & MeV\\
m_{\tau}  & = & 1777.1   &  & \displaystyle{^{+0.4}_{-0.5}} & MeV
\ea
\ee
one finds
\be
\vartheta_{ML} = 45.0003^{\circ}\,\pm\,0.0012^{\circ}\,\,\,,
\ee
which shows the excellent agreement with experiments of the mass formula
proposed by Koide. 

Thus, we propose to extend these considerations to the masses of the 
quarks. Introducing for the {\it down} quarks the 
vector {\bf D}=$(\sqrt{m_d},\sqrt{m_s},\sqrt{m_b})$, 
and using for the quark masses the numerical values most recently 
evaluated {\it via} QCD sum rules \cite{DomBPD}%
\footnote{Similar results should be obtained using the ``experimental'' 
ranges for quark masses reported in \cite{PDB}. We choose to use the 
values estimated in \cite{DomBPD} for their relatively small errors.}
\be
\ba{lclcll}
m_d  & = & 8.3 & \pm & 2.9 & MeV\\
m_s  & = & 175  & \pm & 25 & MeV\\
m_b  & = & 4700 & \pm & 70 & MeV\,\,,
\ea
\ee
the angle formed by the {\bf M} and {\bf D} is
\be
\vartheta_{MD} = 45.6^{\circ}\,\pm\,0.6^{\circ}\,\,\,
\ee
very close to $\vartheta_{ML}$. It is worth noting
that {\bf L} and {\bf D} are not collinear and not coplanar with {\bf 
M}. The cosine of the angle formed by the vector 
{\bf D}$\,\times\,${\bf M} and {\bf L} is equal to $0.70\,\pm\,0.02$ 
excluding definitely coplanarity.

If one applies the same procedure to
the {\it up} quark masses \cite{DomBPD}  
\be
\ba{lclcll}
m_u  & = & 3.7 & \pm & 2.8  & MeV\\
m_c  & = & 1460 & \pm & 70 & MeV  \\
m_t  & = & 174  & \pm & 16 & GeV\,\,,
\ea
\ee
and defines the vector 
{\bf U}=$(\sqrt{m_u},\sqrt{m_c},\sqrt{m_t})$, one obtains
\be
\vartheta_{MU} = 50.9^{\circ}\,\pm\,0.2^{\circ}\,\,\,.
\ee
It is easy to verify that, with the quoted values for the charged 
fermion masses, the vectors {\bf M}, {\bf L}, {\bf D}, {\bf U} do not show 
notable properties of coplanarity among them.

Now assuming for Dirac neutrino masses the same property of {\it up}
quarks, {\it i.e.} that the vector $(\sqrt{m_{\nu_e}},\sqrt{m_{\nu_{\mu}}},
\sqrt{m_{\nu_{\tau}}})$ forms an angle of
$50.9^{\circ}\,\pm\,0.2^{\circ}$ 
with {\bf M}, and making the reasonable assumption that $m_{\nu_e}$ is much
smaller than the others, and therefore to neglect it leads to an angle
slightly larger than $\vartheta_{MN}$, one gets: 
\be 
\frac{1}{\sqrt{3}} \,\frac{
\sqrt{m_{\nu_{\mu}}} + \sqrt{m_{\nu_{\tau}}} }
{\sqrt{m_{\nu_{\mu}} + m_{\nu_{\tau}}}  } \,\,\leq\,\, 0.63 \,\pm\, 0.16
\label{e:dis}
\ee
and a ratio
\be
\frac{m_{\nu_{\mu}}}{m_{\nu_{\tau}}} \,\leq \,(9.6\,\pm\,0.6)\times\,10^{-3}
\label{e:ratio}
\ee
for the Dirac masses of the two neutrinos.

\begin{figure}[tbp]
\epsfysize=18cm
\epsfxsize=16cm
\epsffile{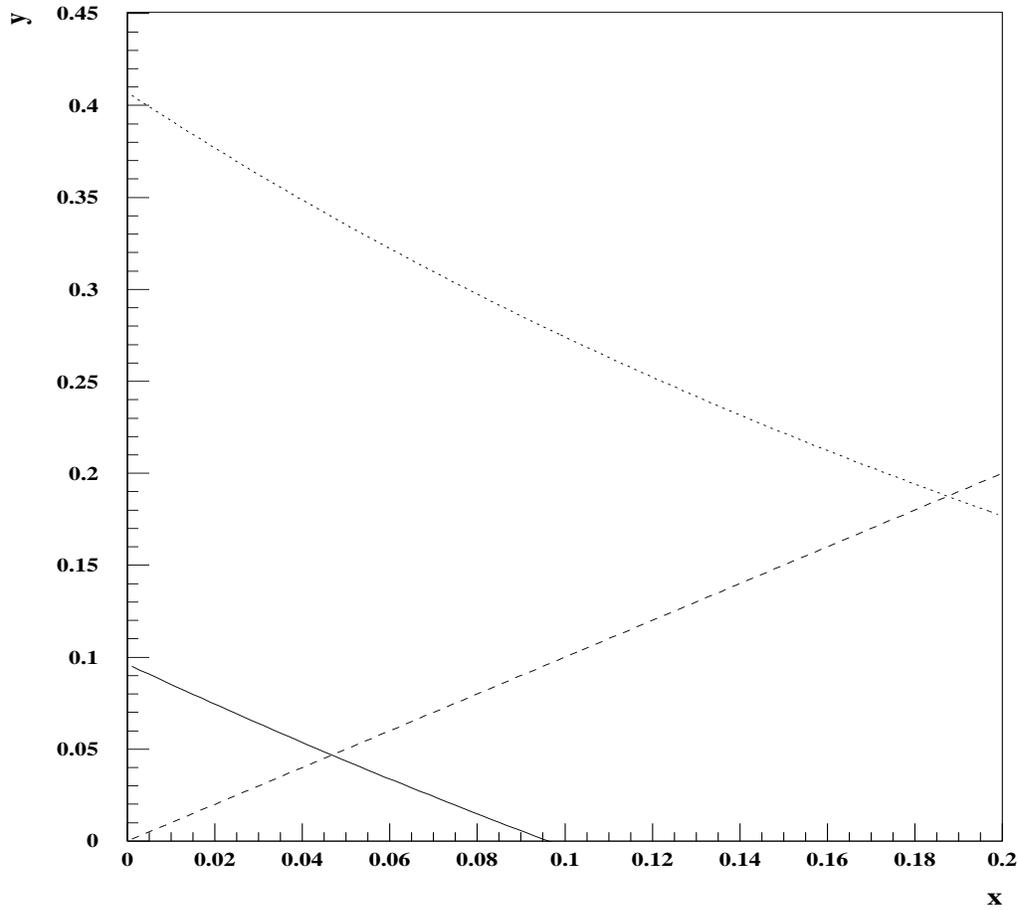}
\caption[]{{\it 
The allowed (1-$\sigma$) range for the ratio 
$y=\sqrt{ m_{\nu_{\mu}}/m_{\nu_{\tau}} }$ versus 
$x=\sqrt{ m_{\nu_{e}}/m_{\nu_{\tau}} }$ is reported. The continuous 
line corresponds to the central value in equation (\ref{e:dis1}),
the dotted one to the 1-$\sigma$ upper limit for $y(x)$. The 
dashed line is the plot for the equation $y=x$.}}
\protect\label{fig:range}
\end{figure}

By fixing the mass of tau neutrino to the cosmologically relevant value, 
$m_{\nu_{\tau}}\sim 7\,eV$ \cite{Bere}, we get from (\ref{e:ratio})
\be
m_{\nu_{\mu}} \leq \,(6.7\,\pm\,0.4)\times\,10^{-2}\,eV\,.
\ee
Note that this upper limit is consistent with the value 
$2.4\,\cdot\,10^{-3}\,eV$ \cite{Ansespo} which would enable one 
to solve the solar neutrino problem in terms of MSW mechanism \cite{MSW}.\\
Now, relaxing the assumption that $m_{\nu_e} = 0$, the equation 
(\ref{e:dis}) becomes:
\be 
\frac{1}{\sqrt{3}} \,\frac{
\sqrt{m_{\nu_{e}}} + \sqrt{m_{\nu_{\mu}}} + \sqrt{m_{\nu_{\tau}}} }
{ \sqrt{m_{\nu_{e}} + m_{\nu_{\mu}} + m_{\nu_{\tau}} }  
} \,\,=\,\, 0.63 \,\pm\, 0.16\,.
\label{e:dis1}
\ee
In Figure \ref{fig:range} we plot the (1-$\sigma$) allowed region for 
the ratio 
$y\equiv\sqrt{ m_{\nu_{\mu}}/m_{\nu_{\tau}} }$ versus 
$x\equiv\sqrt{ m_{\nu_{e}}/m_{\nu_{\tau}} }$.
The straight line corresponds to the $y=x$ equation; we assume, in 
fact, for the neutrino masses the same hierarchy of all other fermions
({\it i.e.} $x\,\leq\, y$).

Last, an important feature of our approach to the mass
problem is related to the mass-scale independence of the geometric
structure outlined above. So far, we have used, for the fermion 
masses, the values given in \cite{DomBPD}; 
these values are evaluated at the corresponding typical 
mass scale.\footnote{For light quarks the renormalization 
mass scale is fixed to about $1\,GeV$.}
Now, the geometric structure and the values of the angles 
between mass vectors are unaltered if we choose only one renormalization 
scale ($\mu$) for the fermion masses. For example, choosing $\mu = 1\,GeV$, 
one can simply verify, using the results about running masses 
given in \cite{KoideR}, that all the numerical values of the cosine of
the angles are unchanged.
Moreover, if we restrict ourselves to consider the mass running below
the $SU(2)\otimes U(1)$ symmetry-breaking scale, the expression
\be 
{\cal F}(\mu) = \frac{\sqrt{m_1M_1} + \sqrt{m_2M_2} + \sqrt{m_3M_3}}
{\sqrt{m_1+m_2+m_3} \sqrt{M_1+M_2+M_3}}\, ,
\ee
which corresponds to the cosine of the angle between mass vectors,
is $\mu$-independent, since the running-mass equation can always 
be written in the diagonal form
\be
\mu \, \frac{d}{d \mu} m_f(\mu) \; = \; - \gamma \, m_f(\mu) \, .
\ee
Remarkably, as the energy scale varies, the angles between the
vectors of such a space remain unaffected by radiative corrections.
This property holds only below the electroweak breaking scale; 
by contrast, above this scale, the renormalization-group equations 
for the masses are coupled and nonlinearities occur \cite{GRZA}. 
The resulting renormalization scale invariance of ${\cal F}(\mu)$ breaks down.

In this paper we have proposed a geometric description for all fermion 
masses, extending the observation made by Foot about charged leptons;
furthermore, this framework enables one to constrain neutrino mass ratios. 
We have shown that this approach is renormalization mass-scale 
independent, and this property proves its geometric nature.

\vspace{1.0truecm}

\noindent
{\bf Acknowledgments}\\
\noindent
We wish to express our thanks to Prof. F. Buccella for several 
discussions and for his unfailing encouragement throughout the course 
of this work. We thank also Dr. G. Esposito for his careful
reading of the manuscript.  

\end{document}